\newcommand{\bra}[1]{\langle {#1} |}
\newcommand{\ket}[1]{| {#1} \rangle }
\newcommand{\braket}[2]{\langle {#1} | {#2} \rangle }

\newcommand{\half}{{\frac{1}{2}}}

\documentclass[letterpaper]{jpconf}
\usepackage{graphicx}
\begin{document}

\title{The effectiveness of quantum operations for eavesdropping on sealed messages}         
\author{Paul A Lopata$^\dagger$ and Thomas B Bahder$^{*}$\footnote{Previous address: Army Research Laboratory, Adelphi, MD}}        
\address{$^\dagger$Sensors and Electronic Devices Directorate, Army Research Laboratory, 2800~Powder~Mill~Road, Adelphi, Maryland, 20783 USA}
\address{$^*$Charles~M~Bowden Research Facility, Aviation and Missile Research, Development and Engineering Center, US Army RDECOM, Redstone Arsenal, Alabama, 35898 USA}
\ead{plopata@arl.army.mil}

\begin{abstract}
A quantum protocol is described which enables a user to send sealed messages and that allows for the detection of active eavesdroppers.  We examine a class of eavesdropping strategies, those that make use of quantum operations, and we determine the information gain versus disturbance caused by these strategies.  We demonstrate this tradeoff with an example and we compare this protocol to quantum key distribution, quantum direct communication, and quantum seal protocols.  
\end{abstract}

\section{Introduction}
We have all become accustomed to sending messages electronically, whether by fax machine, telephone, computer or other electronic media.  Most of these messages contain data that is already publicly known or at least easily found.  Other messages are things we would like to keep to ourselves, and it would be inconvenient if some third party came across the message.  Still other messages are extremely private and resources, jobs, or even lives(!) might be lost if the message fell into the wrong hands.  A great deal of effort is employed to encrypt the messages that fall in this last category, sending them with some sort of code in order to prevent any third party from understanding them even if the messages are intercepted.\cite{Brassard88}  

However, when a message is sent electronically there is no commonly available technology to determine if someone has been trying to intercept the message.  
When sending typed letters, such a technology does exist, albeit in an imperfect form.  We often seal our letters in envelopes.  These envelopes are not secure, that is, they do not \emph{prevent} anyone from opening the envelope and reading the letter inside.  However, when an envelope is received intact, without any tears or other indication that it has been tampered with, we have a strong reason to believe that the message inside has not been seen by anyone since the earlier time when the sender sealed it.   Yet a seal on an envelope is not to be wholly trusted for this task of detecting eavesdroppers.   A skilled person might be able to examine the contents of the sealed envelope in any number of ways: by using x-rays or other similar non-destructive testing methods, by steaming the seal off and re-sealing, or by ripping open the envelope and then placing the letter in a new, forged envelope that matches the original in every detail.

In this paper we introduce a quantum cryptographic protocol that allows two users to send and receive a message in a manner that is, in effect, quite similar to the use of a sealed envelope.  The receiver of the message has the opportunity to check if there have been any active eavesdroppers trying to learn the contents of the message.  And similar to a message sealed in an envelope, the message remains unknown to anyone who is not actively trying to learn the contents.  This protocol has the advantage over sealing letters in envelopes because the limited types of interactions allowed by quantum mechanics prevent someone from eavesdropping on the message without leaving signs of the eavesdropping activities.  

It is important to make it clear that any messages sent using the protocol introduced here are not secure.  That is, an eavesdropper can always choose to take some action in order to determine the content of a message sent using this protocol.  (We give an example of one such effective eavesdropping strategy below.)  The quality that makes this protocol distinct from other methods of message transmission is that any such active eavesdropping strategies will cause an appreciable amount of ``noise'' that is detectable by the message receiver.  The analysis that a message receiver undertakes to place a bound on what an eavesdropper could have learned during a particular message transmission is not undertaken here.  This analysis can be found elsewhere.\cite{Lopata06}  

The goal of this manuscript is to examine a certain class of strategies for eavesdropping on these sealed messages, and it is divided into four parts:  First, the quantum message sealing protocol is introduced.  Following this, we examine a certain class of eavesdropping strategies and describe what an eavesdropper expects to learn by employing such strategies.  Next, we describe the type and amount of disturbance the eavesdropper will cause by such an activity and work out the details of an example from this class of eavesdropping strategies.  We conclude with a discussion of this protocol and its similarities and differences to other quantum cryptographic protocols.   

\section{Message sealing protocol}

We describe the protocol where the message sender named Alice transfers a message to the receiver named Bob.  This message will be a single bit $b$ which is either zero or one.  The protocol utilizes a single quantum mechanical system which has two degrees of freedom.  The standard notation for such a system is used, with $\ket{0}$ and $\ket{1}$ representing vectors that form an orthonormal basis.  The protocol also involves a number of announcements made by the message sender.  These announcements are to be considered as public announcements to which everyone is assumed to have access.

A process, referred to as a single \textit{shot}, will be repeated many times and goes as follows: \\[\parskip]
\textit{Step 1 -} Bob prepares a quantum system, which we refer to as a particle, in one of four pure states: $\ket{0}$, $\ket{1}$, $\ket{+}\equiv (\ket{0}+ \ket{1})/\sqrt{2}$, or  $\ket{-}\equiv(\ket{0}- \ket{1})/\sqrt{2}$.   The decision as to which state to prepare is made at random with equal probability for each state.  He records the state he has prepared and then he sends the particle to Alice.\\
\textit{Step 2 -} Alice makes one of two measurements with equal probability.  She either makes a measurement corresponding to $\sigma_1 = \ket{+}\bra{+} - \ket{-}\bra{-}$ or she makes a measurement corresponding to $\sigma_3 = \ket{0}\bra{0} - \ket{1}\bra{1}$.  Each of these two measurements can be said to have a result $m$ that is either $m=+1$ or $m=-1$.  \\
\textit{Step 3 -} Alice announces whether her measurement corresponded to $\sigma_1$ or $\sigma_3$.  \\
\textit{Step 4 -} Alice makes one of two possible announcements.  With probability $p_a$ she makes a \textit{bit-announcement} (described immediately below) and with probability $(1-p_a)$ she makes a \textit{result-announcement}.  She also makes it known which of the two types of announcement she is making. \\
$\phantom{m}$\textit{Bit-Announcement:}  She announces a bit $c$ that is determined by using the message bit $b$ and the measurement result.  If her measurement yielded the result $m=+1$ then her announced bit $c$ will be the same as the message bit $b$ and if her measurement yielded the result $m=-1$ her announced bit $c$ will be the opposite of the message bit $b$.  \\
$\phantom{m}$\textit{Result-Announcement:} She announces the result of her measurement, $m=+1$ or $m=-1$.

When Bob prepares the particle in the state $\ket{0}$ or $\ket{1}$ and Alice makes a $\sigma_3$ measurement, or when Bob prepares the particle in the state $\ket{+}$ or $\ket{-}$ and Alice makes a $\sigma_1$ measurement we say that Alice's measurement and Bob's state preparation have a matching basis.   They will have a matching basis on half the shots performed.  When this occurs then Bob knows the result of the measurement without Alice having to announce it, provided that the state of the particle did not change from when Bob prepared it to when Alice makes the measurement.   The correlations between Bob's state preparation and Alice's measurement results allow Bob to both determine the message bit and check the channel for any disturbances.  

When Alice makes a measurement in the basis matching Bob's state preparation, Bob determines the message by applying a controlled-bit-flip operation on the announced bit.  When the state in which he prepared the particle is either $\ket{0}$ or $\ket{+}$ then the message bit $b$ is the same as the announced bit $c$ and if he prepared $\ket{1}$ or $\ket{-}$ then the message $b$ is the opposite of the announced bit $c$.  

From an eavesdropper's point of view, the probability that the message bit is one value or the  other is determined from the coded bit-announcements.   When both values of the measurement result are equally likely then both values of the message bit are equally likely (for either bit-announcement).  The four possible initial states that Bob prepares and the two possible measurements were chosen so that either measurement result is equally likely.  Moreover, the only opportunity that an eavesdropper has to change these probabilities is to change the state of the particle when it is traveling from Bob to Alice.  The rules of quantum mechanics allow for the state of a quantum mechanical system to change in two different ways: by a unitary evolution or by a measurement.  If we want to describe the effects of coupling the quantum system composed of the particle to another (auxiliary) quantum system and then letting the state of whole system (particle plus auxiliary) change via unitary evolution of measurement, the entire process can be described as a quantum operation or a generalized measurement on the state of the particle subsystem.\cite{Nielsen00}

In the following sections we examine the case of when an eavesdropper chooses to change the state of the particle by applying a quantum operation.  It is worthwhile to emphasize that while using this type of eavesdropping activity is not optimal,\cite{Lopata06} it provides us with some intuition as to how this protocol can be expected to work.

\section{Information gain from quantum operations}

In this section we quantify what an eavesdropper learns by applying a quantum operation to change the state of the particle as it travels from Bob to Alice.  We first describe quantum operations\cite{Nielsen00} and then tackle the problem of quantifying an eavesdropper's gain by using the Shannon mutual information.\cite{Shannon93} 

A quantum operation $\mathcal E$ acting on states in Hilbert space $\mathcal H$ is described by a set of operators $\{ E_1, \ldots, E_n \}$  subject to the requirement that $\sum_i E_i^\dagger E_i = I$ where $I$ is the identity operator acting on $\mathcal H$.  We say that the quantum operation $\mathcal E$ maps the initial state $\rho$ to final state $\mathcal E(\rho) = \sum E_i \rho E_i^\dagger$.  
A quantum operation is a convex linear map on the space of mixed states, which is to say that if $\rho = p \rho_1 + (1 - p) \rho_2$ with $0 \leq p \leq 1$, then $\mathcal E (\rho) = p \mathcal E (\rho_1) + (1 - p) \mathcal E (\rho_2)$.  A special class of quantum operations are the \textit{unital} quantum operations that map the chaotic state, which is $\frac{1}{d}I$ where $d = \dim (\mathcal H)$, to itself.   

We quantify the amount an eavesdropper learns by using the Shannon mutual information between two random variables: the random variable $B$ which describes the possible values of the message and their probabilities, and the random variable $C$ which describes the possible strings of  bit-announcements and their probabilities.  These strings result from the fact that there will be $N$
shots, and an announcement will be made on each shot.  On some of the shots only the result of the measurement will be announced, and this result does not depend on the message in any way.  Therefore, only the bit-announcements will be of any concern to us in quantifying what the eavesdropper learns.  

The possible messages are $b=0$ and $b=1$ with one-half prior probability each.  

On each shot there are four possible bit-announcements --- $(\sigma_1, 0)$, $(\sigma_1, 1)$, $(\sigma_3, 0)$, and $(\sigma_3, 1)$ --- and when $N$ shots are made, $k$ of which result in bit-announcements (where $0 \leq k \leq N$), there are $4^k$ possible bit-announcement strings.  Because of the probabilistic nature of the protocol, the number of bit-announcements is not fixed.  The probability $p_k$ of making $k$ bit-announcements is found using the binomial distribution
\begin{displaymath}
p_k = {N \choose k} p_a^{\phantom{i}k} (1-p_a)^{N-k} \ .
\end{displaymath} 
We use the symbol $\mathbf{c}$ to denote a bit-announcement string, and we use the symbol $C^{(k)}$ to describe the ensemble of all possible bit-announcement strings of length $k$.  

Given that there are $k$ bit-announcements, the Shannon mutual information $I(C^{(k)}:B)$ is calculated using 
\begin{equation}\label{Shannon}
I(C^{(k)}:B)  =  \sum_{\mathbf{c}^{(k)}}\Biggl[\Pr (\mathbf{c}) \log \frac{1}{\Pr (\mathbf{c})}  +  \frac{1}{2}\sum_{b = 0}^1  \Pr(\mathbf{c} \; | \; b) \log \Pr(\mathbf{c} \; | \; b)\Biggr] 
\end{equation}
where the sum over $\mathbf{c}^{(k)}$ indicates that this sum is taken over all $4^k$ bit-announcement strings of length $k$.  This can be used to determine the expected mutual information when taking the weighted sum over the various possible lengths of bit-announcement strings,
\begin{equation} \label{n_dependent_Shannon}
I(C:B)=\sum_{k=0}^N p_k I(C^{(k)}:B) \ .
\end{equation}
This can be calculated once the probabilities $\Pr(\mathbf{c}|b)$ are known for every $\mathbf{c}$ and both values of $b$. The remainder of this section is devoted to determining these probabilities, which will change depending upon which quantum operation is applied.  

For a given value of the message, the probabilities of the four bit-announcements depend upon the probability of Alice getting the $m=+1$ measurement result.  That is, 
\begin{displaymath}
\begin{array}{rcccl}
\Pr ( \sigma_i, c=b | b) & = & \Pr(m=+1| \sigma_i) \Pr(\sigma_i) & = & \Pr(m=+1| \sigma_i)/2 \ , \\
\Pr ( \sigma_i, c\neq b | b) & = & \Pr(m=-1| \sigma_i) \Pr(\sigma_i) & = & \Pr(m=-1| \sigma_i)/2 \ ,
\end{array}
\end{displaymath}
where $i = 1,3$ and $b=0,1$.  The notation $\Pr(m=+1|\sigma_i)$, for example, is used to mean that this is the probability that the result $m=+1$ will be found when a measurement that corresponds to $\sigma_i$ is made on the particle and $\Pr(\sigma_i)$ is the probability that the measurement corresponding to $\sigma_i$ will be performed.  

Of course, the machinery of quantum mechanics requires us to specify the state of the particle in order to calculate a probability of a certain measurement result.  From an eavesdropper's point of view, if she does nothing to the particle then there are four possible states with equal probability.  So $\Pr(m=\pm1| \sigma_i) = \frac{1}{4}(\Tr(\half(I \pm \sigma_i) \ket{0}\bra{0} ) + \Tr(\half(I \pm \sigma_i) \ket{1}\bra{1} ) + \Tr(\half(I \pm \sigma_i) \ket{+}\bra{+} ) + \Tr(\half(I \pm \sigma_i) \ket{-}\bra{-} ))$ where $i=1,3$.  By the linearity of the Trace function, this is equivalent to $\Pr(m=\pm1| \sigma_i) = \Tr(\half(I \pm \sigma_i) \half I )$.   In this way, it is quite reasonable to say that the state of the particle, to the eavesdropper's best description, is the chaotic state $\rho = \half I$.  

When an eavesdropper applies a quantum operation $\mathcal{E}$ to change the state of the particle, it will in general change each of the four possible initial states differently.  By the linearity of the Trace function and the convex linearity of the quantum operation $\mathcal{E}$,  the probability of $m=\pm1$ can be calculated for the state $\rho' = \mathcal{E}(\half I)$.  That is, $\Pr(m=\pm1| \sigma_i) =  \Tr(\half(I \pm \sigma_i) \mathcal{E}(\half I) )$ for $i = 1,3$. 

Every (generally mixed) state of a two-level quantum system can be described by $\rho(\lambda \mathbf{v}) = \half ( I + \lambda [v_1 \sigma_1 + v_2 \sigma_2 + v_3 \sigma_3 ] )$ where $v_1^2 + v_2^2+v_3^2 = 1$, $\sigma_2 = i \sigma_1 \sigma_3$, and $0 \leq \lambda \leq 1$.  This ``Bloch sphere'' description of the two-level state can be pictured as a vector $\lambda \mathbf{v}$ in a real three dimensional space.  When $\mathcal E (\half I)=  \half ( I + \lambda [v_1 \sigma_1 + v_2 \sigma_2 + v_3 \sigma_3 ] )$, the probabilities for the four possible announcements are shown in Table \ref{announcement_probs}.
\begin{table} 
\caption{\label{announcement_probs}The probabilities for the four results relevant to the bit-announcements, given that an eavesdropper acts with a quantum operation $\mathcal{E}_{\lambda \mathbf v}$ that maps the chaotic state to $\rho(\lambda \mathbf{v})$.}
\begin{center}
\begin{tabular}{rcl}
\br
$\Pr(m=+1| \sigma_1, \mathcal E_{\lambda\mathbf v} )$ & $=$ & $\half (1 + \lambda v_1)$\\[0.1cm]
$\Pr(m=-1| \sigma_1, \mathcal E_{\lambda\mathbf v} )$ & $=$ & $\half (1 - \lambda v_1)$\\[0.1cm]
$\Pr(m=+1| \sigma_3, \mathcal E_{\lambda\mathbf v} )$ & $=$ & $\half (1 + \lambda v_3)$\\[0.1cm]
$\Pr(m=-1| \sigma_3, \mathcal E_{\lambda\mathbf v} )$ & $=$ & $\half (1 - \lambda v_3)$\\[0.1cm]
\br
\end{tabular}
\end{center}
\end{table}
If an eavesdropper applies the same quantum operation each time a particle is sent from Bob to Alice, the probabilities for each bit-announcement string is found by taking the product of the probabilities of each of the four announcements, with each of the probabilities appearing in the product the same number of times that that announcement appears in the string.

We can now calculate the mutual information for any quantum operation by calculating the probabilities for each bit-announcement string and then using Equations (\ref{Shannon}) and (\ref{n_dependent_Shannon}). 

To summarize this section, we have described how to calculate the mutual information which quantifies what an eavesdropper expects to learn about the message given a particular quantum operation used as an eavesdropping strategy.  In the next section, we determine the amount of ``noise'' that such eavesdropping strategies cause.

\section{Disturbance caused by quantum operations}

In the previous section we focused on the bit-announcements and ignored the result-announcements.  In this section we will do the opposite.  The bit-announcements are used by both Bob and any eavesdroppers to determine the message, but the result-announcements are of no use to the eavesdropper and serve Bob's purpose to check the channel for ``noise''.

There are sixteen different event statistics that are kept by Bob relating to the measurement-announcements: four possible initial states, two possible measurement types, and two possible measurement results for each measurement.   Out of these sixteen, there are four events that would be the most surprising to Bob, and would each indicate that the state of the particle, when Alice measured it, was not the same as the one he had prepared.  These four types of events will be referred to as \emph{mismatches} and are shown in Table \ref{mismatches}.
\begin{table} 
\caption{\label{mismatches}The four events that correspond to mismatches.}  
\begin{center}
\begin{tabular}{ccc}
\br
Bob prepares the state & Alice measures & measurement result \\
\hline
$\ket{+}$ & $\sigma_1$ & $m = -1$ \\
$\ket{-}$ & $\sigma_1$ & $m = +1$ \\
$\ket{0}$ & $\sigma_3$ & $m = -1$ \\
$\ket{1}$ & $\sigma_3$ & $m = +1$ \\
\br
\end{tabular}
\end{center}
\end{table}
The probability of a mismatch, on a particular shot, is 
\begin{eqnarray} \label{mismatch_eq}
\Pr(\mathrm{mismatch}) &= & \Pr(\ket{+}, \sigma_1, -1 )+\Pr(\ket{-}, \sigma_1, +1 )+\Pr(\ket{0}, \sigma_3, -1 )+\Pr(\ket{1}, \sigma_3, +1 ) \nonumber \\
& = &\frac{1}{4}\Bigl(\Pr(\sigma_1, -1|\ket{+} )+\Pr( \sigma_1, +1 |\ket{-})+\Pr(\sigma_3, -1 |\ket{0})+\Pr(\sigma_3, +1| \ket{1} ) \Bigr) \nonumber \\
& = & \frac{1}{8}\Bigl(\Pr( -1|\ket{+},\sigma_1 )+\Pr(+1 |\ket{-}, \sigma_1)+\Pr( -1 |\ket{0}, \sigma_3)+\Pr( +1| \ket{1}, \sigma_3)\Bigr).
\end{eqnarray}
Of course, when Bob analyzes the data, a mismatch can only occur on a particular shot if the bases are matched up.  A factor of $1/2$ disappears when we account for this to give the probability that there will be a mismatch error on a shot when the bases are matched.  For a fixed quantum operation $\mathcal E$ employed by an eavesdropper, these probabilities are easily calculated.  Note that these probabilities depend upon the final states $\mathcal E(\ket{+}\bra{+})$, $\mathcal E(\ket{-}\bra{-})$, $\mathcal E(\ket{0}\bra{0})$, and $\mathcal E(\ket{1}\bra{1})$, and not just on the evolution of the chaotic state.  In general, there are many different quantum operations that have the same effect on the chaotic state.  (The exception to this is when the chaotic state is mapped to a pure state, in which case it is easily seen by the convex linearity of quantum operations that every initial state must be mapped to that pure state.)

\section{An Example}\label{example}

Let us now examine a family of eavesdropping strategies that utilize the quantum operation $\mathcal{E}_x$, where $x$ is a parameter which falls in the range $0 \leq x \leq 1$.  When $x = 0$ the strategy corresponds to the eavesdropper doing nothing (and as we shall see, learning nothing),  and when $x = 1$ it corresponds to a quantum operation eavesdropping strategy with the greatest mutual information.  

The quantum operation $\mathcal{E}_x$ can be achieved by coupling the initial state $\rho$ (from Bob) to an auxiliary quantum system in the state $\ket{\phi}$, letting the coupled system evolve unitarily (described by some unitary operator $U$ that acts on the combined system) and then tracing over the auxiliary system.   The unitary operator acts as follows:
\begin{eqnarray*}
U \Bigl( \, \ket{0}\otimes \ket{\phi} \, \Bigr) & = & \ket{0}\otimes \ket{F}  \ \  \equiv \ \ \ket{\Gamma_0}\\
U \Bigl( \, \ket{1}\otimes \ket{\phi} \, \Bigr) & = & \sqrt{x} \, \ket{0}\otimes \ket{G} + \sqrt{1-x}\, \ket{1}\otimes \ket{F}  \ \ \equiv \ \ \ket{\Gamma_1} \ ,
\end{eqnarray*}
where $\braket{F}{G}=0$ and $\braket{F}{F}=\braket{G}{G}=1$.  The fact that $\braket{0}{1}\braket{\phi}{\phi} = \braket{\Gamma_0}{\Gamma_1}$ is sufficient to show that such a unitary operator $U$ exists.  The action of the quantum operation $\mathcal{E}_x$ on any initial pure state $\ket{\eta}$ is found by tracing over the auxiliary subsystem after performing the unitary transformation $U$:
\begin{displaymath}
\mathcal{E}_x\Bigl(\ket{\eta}\bra{\eta}\Bigr) = \Tr_{\mathrm{aux}} \biggl( U \Bigl(\ket{\eta}\otimes \ket{\phi}\Bigr)\Bigl(\bra{\eta}\otimes \bra{\phi}\Bigr)U^\dagger   \biggr) \ .
\end{displaymath}
By the convex linearity of quantum operations we also know the action of $\mathcal{E}_x$ on any mixed state as well.  From the preceeding considerations, it is straightforward to show that $\mathcal{E}_x$ acts on the relevant initial states in the following way:
\begin{eqnarray*}
\mathcal{E}_x \Bigl(\ket{0}\bra{0}\Bigr) & = &  \ket{0}\bra{0} \\
\mathcal{E}_x \Bigl(\ket{1}\bra{1}\Bigr) & = &  x \, \ket{0}\bra{0} + (1-x) \, \ket{1}\bra{1}\\
\mathcal{E}_x \Bigl(\ket{+}\bra{+}\Bigr) & = &  \frac{1}{2}\Bigl[ (1+x) \, \ket{0}\bra{0} + (1-x) \, \ket{1}\bra{1} + \sqrt{1-x} \Bigl(\ket{0}\bra{1} + \ket{1}\bra{0}\Bigr) \Bigr]\\
\mathcal{E}_x \Bigl(\ket{-}\bra{-}\Bigr) & = &  \frac{1}{2}\Bigl[ (1+x) \, \ket{0}\bra{0} + (1-x) \, \ket{1}\bra{1} - \sqrt{1-x} \, \Bigl(\ket{0}\bra{1} + \ket{1}\bra{0}\Bigr) \Bigr] \ ,
\end{eqnarray*}
from which it is easy to see that
\begin{displaymath}
\mathcal{E}_x \Bigl( \half I \Bigr)  =   \frac{1}{2}\Bigl[ (1+x) \, \ket{0}\bra{0} + (1-x) \, \ket{1}\bra{1}  \Bigr] = \half (I + x \sigma_3)  .
\end{displaymath}

The probability of a mismatch,  calculated using Equation (\ref{mismatch_eq}), for this quantum operations is $\frac{1}{4}(1+x - \sqrt{1-x})$.  

\begin{table} 
\caption{\label{example_table}The probabilities, from the eavesdropper's point of view, of the four possible bit-announcements for a given value of $b$ when the quantum operation $\mathcal{E}_x$, introduced in Section \ref{example}, is applied.}  
\begin{center}
\begin{tabular}{lcc}
\br
                                  & $\underline{b=0}$ & $\underline{b=1}$ \\
$\Pr(\sigma_1, c=0 | b) \ \ \ \ $ & $\ 1/4 \ $ & $\ 1/4 \ $ \\
$\Pr(\sigma_1, c=1 | b) \ \ \ \ $ & $\ 1/4 \ $ & $\ 1/4 \ $ \\
$\Pr(\sigma_3, c=0 | b) \ \ \ \ $ & $\ (1+x)/4 \ $ & $\ (1-x)/4 \ $ \\
$\Pr(\sigma_3, c=1 | b) \ \ \ \ $ & $\ (1-x)/4 \ $ & $\ (1+x)/4 \ $ \\
\br
\end{tabular}
\end{center}
\end{table}
In order to calculate the mutual information for this quantum operation, we must be able to determine the values of $\Pr(\mathbf{c}|b, \mathcal{E}_x)$, that is, the probability of a every string of result-announcements $\mathbf{c}$ given each value of $b$.  If a particular string of $k$ result-announcements $\mathbf{c}(k, d_1, d_2, d_3, d_4)$ consists of $(\sigma_1, c\!=\!0 )$ announced $d_1$ times, $(\sigma_1, c\!=\!1 )$ announced $d_2$ times, $(\sigma_3, c\!=\!0 )$ announced $d_3$ times, and $(\sigma_3, c\!=\!1)$ announced $d_4$ times --- in any order --- then the probability for this announcement to occur is 
\begin{eqnarray*}
\Pr\Bigl(\mathbf{c}(k, d_1, d_2, d_3, d_4)| b\!=\!0, \mathcal{E}_x\Bigr) & = & \left(\frac{1}{4}\right)^{k}(1+x)^{d_3}(1-x)^{d_4} \equiv p_{x, k, d_3, d_4}\\
\Pr\Bigl(\mathbf{c}(k, d_1, d_2, d_3, d_4)| b\!=\!1, \mathcal{E}_x\Bigr) & = & \left(\frac{1}{4}\right)^{k}(1-x)^{d_3}(1+x)^{d_4} \equiv q_{x, k, d_3, d_4} \ .
\end{eqnarray*}
This calculation utilizes the probabilities for the single announcements found in Table \ref{example_table}.  There are $k! / (d_3! d_4! (k-d_3-d_4)!)$ different strings of $k$ bit announcements that share this same probability (for each value of $b$).

Using these results, we can now calculate the mutual information.  
\begin{eqnarray}
I(C^{(k)}:B) &=&- \sum_{d_3 = 0}^k \sum_{d_4=0}^{k-d_3}  \frac{ k!}{d_3! d_4! (k\!-\!d_3\!-\!d_4)!}  \Biggl[ \frac{1}{2}\Bigl(  p_{x, k, d, d_3} + q_{x, k, d, d_3} \Bigr) \log  \frac{1}{2} \Bigl(  p_{x, k, d, d_3} + q_{x, k, d, d_3} \Bigr)  \nonumber \\
& & \label{SpecificMutualInfo}\ \ \ \ \ \ \ \ \ \ \ \ \ \ \ \ \ \ \ -\frac{1}{2}  \Bigl(  p_{x, k, d, d_3} \log p_{x, k, d, d_3} - q_{x, k, d, d_3}\log q_{x, k, d, d_3} \Bigr) \Biggr]
\end{eqnarray}

\begin{figure} 
\begin{minipage}[t]{18pc}
\includegraphics[width=18pc]{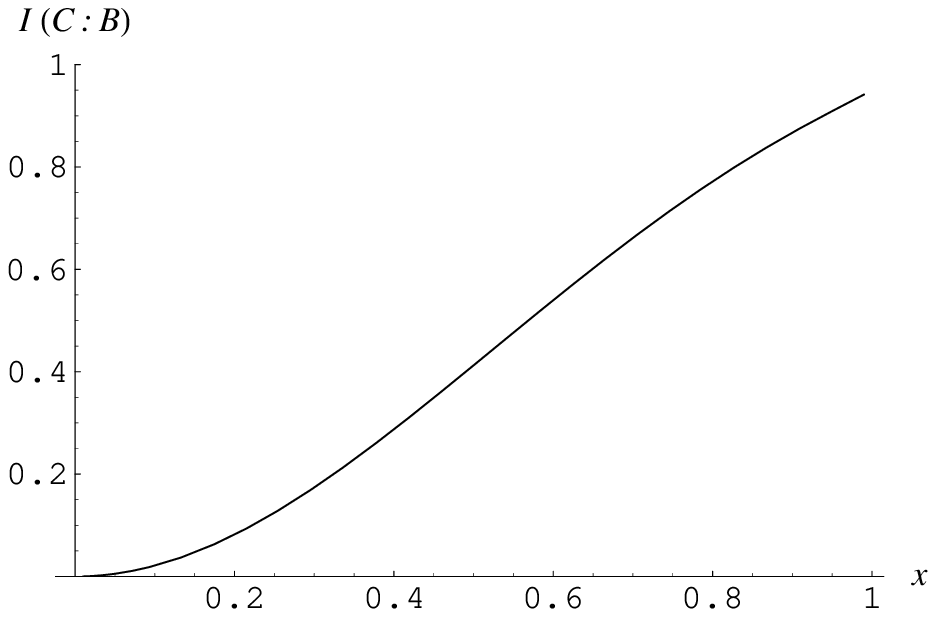}
\caption{\label{MutualInfo}Mutual information as a function of $x$, describing the amount an eavesdropper learns about the message bit given that she uses the quantum operation $\mathcal{E}_x$ on each shot when Bob sends $N=119$ particles and Alice has probability $p_a = 0.01$ of making a bit-announcement.}
\end{minipage}\hspace{2pc}%
\begin{minipage}[t]{18pc}
\includegraphics[width=18pc]{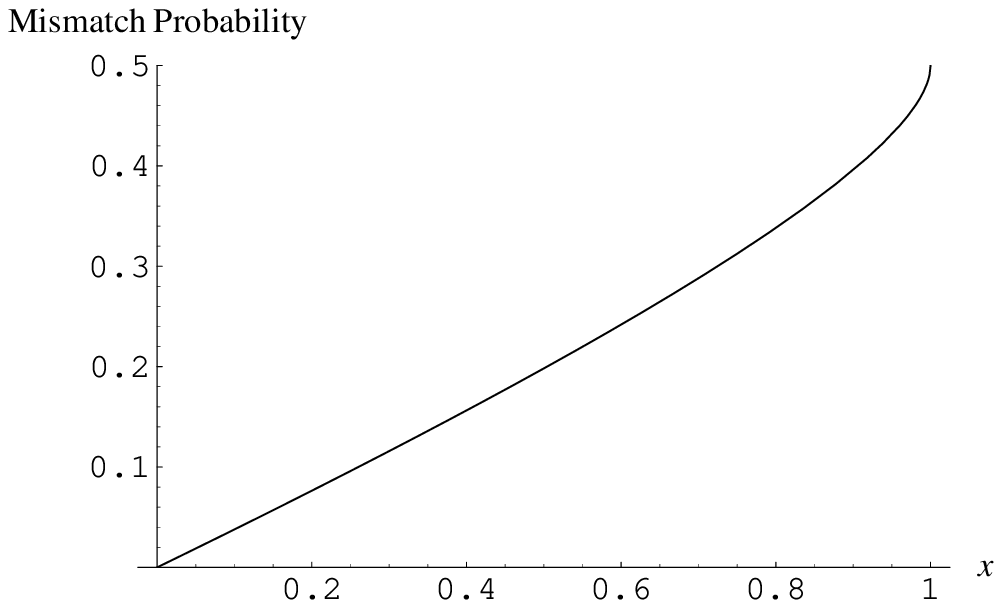}
\caption{\label{mismatch}Probability of a mismatch as a function of $x$ when an eavesdropper uses the quantum operation $\mathcal{E}_x$ on each shot.}
\end{minipage} 
\end{figure}

If we choose some exemplary values of $p_a$ and $N$, this will give us some numerical results for the mutual information.  Say that Alice sets $p_a = 0.05$ and Bob agrees with Alice to send $N=119$ particles in order to communicate the value of a single bit.  This choice of $p_a$ and $N$ gives them slightly more than a $95\%$ chance of matching their bases on a shot when a result-announcement is made.  The mutual information $I(C:B)$, when an eavesdropper applies the quantum operation $\mathcal{E}_x$ on every shot, is plotted for all values of $0 \leq x \leq 1$ in Figure \ref{MutualInfo}.  Compare this with the disturbance caused, quantified by the probability of a mismatch, by applying the same quantum operation, which is shown in Figure \ref{mismatch}.

As a final note, this example demonstrates that a passive eavesdropper learns nothing about the message.  That is, if we describe a passive eavesdropper as someone who is only listening to the announcements that Alice makes but does not interfere with the particles in any way,\cite{Brassard88} that person's eavesdropping strategy would correspond to $\mathcal{E}_x$ when $x=0$.  It is easily seen from the Figures that this strategy causes the eavesdropper to learn nothing and also to causes no disturbance.  

\section{Discussion}

This protocol represents  something new in the field of cryptography.  It provides the message receiver with a way to check if an eavesdropper is attempting to access the message.  The analysis shown here demonstrates both the amount learned by an eavesdropper and the disturbance caused, measured in the number of mismatches, when an eavesdropper employs a particular quantum operation.  

As shown in the example above, this protocol is not secure against active attacks in which an eavesdropper interacts with the particles as they travel from the message receiver to the message sender. However, this example also demonstrates that such attacks cause a disturbance in the system, which can be quantified by the number of mismatches found by the message receiver.  A more general analysis a message receiver's bound on the amount of information an eavesdropper could have learned during a particular transmission is taken up elsewhere.\cite{Lopata06}  

The protocol discussed here has similarities to other quantum cryptography protocols that have been introduced and it is worthwhile to examine these similarities, as well as what makes this current protocol distinct.  The three types of quantum cryptographic protocols that will be discussed here are quantum key distribution (QKD) protocols, quantum secure direct  communication (QSDC) protocols, and quantum seal protocols.  

The main distinction between this new protocol and the QKD protocols is that the goal of QKD is to develop a shared private key between two parties while here it is important that a particular message gets transmitted.  Said in a different way, each party in a QKD setting starts with nothing and ends up with a random string of bits, but neither one of them cares which string of bits results from the process, so long as they share the same one.  Here, one party starts with a particular string of bits --- the message --- and when the process ends the other party will (hopefully) have the message as well.  (There is a tunably small probability that the process will be unsuccessful.\cite{Lopata06})  Of course, in QKD the random string of bits can later be used to encrypt a message (which can be sent on a classical channel), but the QKD process itself transfers no information.

It is worthwhile mentioning that this current protocol is very similar, in some ways, to a specific QKD protocol, called BB84.\cite{Bennett84}  The two protocols use the same four initial possible states and the same two measurements.  The difference between the two is the classical messages that are sent and how these messages are used.  These two protocols are so similar that if two users have a system that implements BB84 then they should be able to implement this new protocol with only minor modifications to the system.  

The second type of quantum cryptographic protocol that we will discuss is the so-called ``quantum secure direct communication'' (QSDC).\cite{Bostrom02}     The greatest similarity between the QSDC protocols and the one introduced here is that they both use quantum states of some transferred system to transmit a message from one party to another, rather than generating a key.  Moreover, this is done without the use of any pre-shared key.  However, the goal of QSDC is to transmit the messages securely (that is, to prevent any eavesdropper from understanding the message), while the goal of the protocol introduced here is to detect the activity of any active eavesdroppers.  

The final comparison we will make is with those quantum cryptographic protocols that have been called ``quantum seal'' protocols.\cite{Bechmann-Pasquinucci03}  These quantum seal protocols are distinct from the current one.   The goal of the quantum seal protocols is for a message sender to prepare a quantum mechanical system in some initial state so that someone else can determine the message by making a measurement on that quantum mechanical system.  Moreover, the message preparer also creates correlations between the quantum mechanical system and a second quantum mechanical system so that a measurement can be made, by the message preparer, on the second system to determine if anyone has read the message.  The major distinction between these quantum seal protocols and the protocol introduced here is that protocol introduced here has a preferred message receiver (the person who sends the particles to the message sender) who can check if anyone else has tried to read the message, while in these earlier quantum seal protocols\cite{Bechmann-Pasquinucci03} all receivers are on equal footing and it is the message sender who can check if someone has accessed the message.

We conclude this discussion by emphasizing that the protocol introduced here is neither a QKD protocol, nor a QSDC protocol, nor a quantum seal protocol.  It has distinct goals and the various security (or no-security) proofs that have been applied to these earlier protocols do not apply here.  

\ack
This work was funded in part by the Disruptive Technology Office (DTO) and by the Army Research Office (ARO). This research was performed while Paul Lopata held a National Research Council Research Associateship Award at the Army Research Laboratory.


\section*{References}
\begin{thebibliography}{20}
\bibitem{Brassard88} Brassard G \textit{Modern Cryptology} 1988 (Spring-Verlag New York, Inc.)
\bibitem{Lopata06} Lopata P and Bahder T, manuscript in preparation
\bibitem{Nielsen00} Nielsen M and Chuang I 2000 \textit{Quantum Computation and Quantum Information} (Cambridge University Press)
\bibitem{Shannon93} Shannon C 1993 \textit{Claude Elwood Shannon Collected Papers} (IEEE Press) p 84
\bibitem{Bennett84} Bennett C and Brassard G 1984 \textit{Proceedings of IEEE International Conference on Computers, Systems and Signal Processing} (IEEE Press) pp 175--179
\bibitem{Bostrom02} Bostr\"om K and Felbinger T 2002 \textit{Physical Review Letters} \textbf{89} 187902
\nonum W\'ojcik A 2003 \textit{Physical Review Letters} \textbf{90} 157901
\nonum Deng F-G, Long G L, and Liu X-S 2003 \textit{Physical Review A} \textbf{68} 042317
\nonum Deng F-G and Long G L 2004 \textit{Physical Review A} \textbf{69} 052319
\nonum Lucamarini M and Mancini S \textit{Physical Review Letters} \textbf{94} 140501 
\nonum and others.
\bibitem{Bechmann-Pasquinucci03} Bechmann-Pasquinucci H 2003 Quantum Seals \textit{Preprint} quant-ph/0303173
\nonum  Bechmann-Pasquinucci H, D'Ariano G M, and Macchiavello C 2005 Impossibility of Perfect Sealing of Classical Information \textit{Preprint} quant-ph/0501073
\nonum Singh S K and Srikanth R 2005 \textit{Physica Scripta} \textbf{71} pp 433--5 
\nonum He G-P 2005 \textit{Physical Review A} \textbf{71} 054304
\nonum Chau H F 2006 \textit{Physics Letters A} \textbf{354} pp 31--4
\end{thebibliography}
\end{document}